\documentclass{emulateapj}

\usepackage{multirow}

\begin{document}

\shorttitle{SMC Cepheid P-L Relations}
\shortauthors{Ngeow et al.}

\title{Period-Luminosity Relations Derived From the OGLE-III Fundamental Mode Cepheids II: The Small Magellanic Cloud Cepheids}
\author{Chow-Choong Ngeow\altaffilmark{1}, Shashi M. Kanbur\altaffilmark{2}, Anupam Bhardwaj\altaffilmark{3} and Harinder P. Singh\altaffilmark{3}}
\altaffiltext{1}{Graduate Institute of Astronomy, National Central University, Jhongli 32001, Taiwan}
\altaffiltext{2}{Department of Physics, SUNY Oswego, Oswego, NY 13126, USA}
\altaffiltext{3}{Department of Physics \& Astrophysics, University of Delhi, Delhi 110007, India}

\begin{abstract}

In this paper we present multi-band period-luminosity (P-L) relations for fundamental mode Cepheids in the SMC. The optical $VI$-band mean magnitudes for these SMC Cepheids were taken from the third phase of the Optical Gravitational Lensing Experiment (OGLE-III) catalog. We also matched the OGLE-III SMC Cepheids to 2MASS and SAGE-SMC catalog to derive mean magnitudes in the $JHK$-bands and the four {\it Spitzer} IRAC bands, respectively. All photometry was corrected for extinction by adopting the Zaritsky's extinction map. Cepheids with periods smaller than $\sim2.5$~days were removed from the sample. In addition to the extinction corrected P-L relations in nine filters from optical to infrared, we also derived the extinction-free Wesenheit function for these Cepheids. We tested the nonlinearity of these SMC P-L relations (except the $8.0\mu\mathrm{m}$-band P-L relation) at 10~days: none of the P-L relations show statistically significant evidence of nonlinearity. When compared to the P-L relations in the LMC, the $t$-test results revealed that there is a difference between the SMC/LMC P-L slopes only in the $V$- and $J$-band. Further, we found excellent agreement between the SMC/LMC Wesenheit P-L slope. The difference in LMC and SMC Period-Wesenheit relation LMC and SMC zero points was found to be $\Delta \mu=0.483\pm0.015$~mag. This amounts to a difference in distance modulus between the LMC and SMC.

\end{abstract}


\section{Introduction}

The period-luminosity relation (the Leavitt Law, hereafter P-L relation) for Small Magellanic Cloud (SMC) Cepheids was first presented in \citet{lea12}. Since then, the SMC P-L relations have been derived from optical to near-infrared \citep[for examples, see][and references therein]{arp60,pay65,way84,wel84,vis85,wel85,cal86,lan86,mat86,wel87,cal91,smi92,lan94,nem94,sha02} based on relatively small number of SMC Cepheids. In 1999, the second phase of the Optical Gravitational Lensing Experiment (hereafter OGLE-II) released a catalog that contained more than $450$ fundamental mode Cepheids located in $\sim2.4$ square degree at the center of SMC \citep{uda99b}. A number of SMC P-L relations have been derived in literature based on these OGLE-II SMC Cepheids \citep{uda99a, gro00, sto04, tam08, san09, bon10}. Independently, the EROS (Exp\'{e}rience de Recherche d'Objets Sombres) Collaboration also derived the P-L relations based on a large number of SMC Cepheids in customized filters \citep{sas97,bau99,mar99}. 

In 2010, a catalog for an even larger number of Cepheids in SMC was released from the the third phase of OGLE operation \citep[hereafter OGLE-III, see][]{sos10}. Compared to OGLE-II, more than $2600$ fundamental mode SMC Cepheids were included in the OGLE-III SMC catalog as a result of larger survey area \citep{sos10}. Although the $VI$-band SMC P-L relations were also derived in \citet{sos10}, these P-L relations were not corrected for extinction. Subsequently, \citet{sub14} used the OGLE-III $VI$-band photometry for Cepheids to investigate the spatial structure of SMC. In terms of the near infrared $JHK$-band P-L relations, \citet{mat11} presented preliminary SMC P-L relations in three period bins by matching the OGLE-III SMC Cepheids to the single-epoch Magellanic Clouds point source catalogs \citep{kat07} based on the Infrared Survey Facility (IRSF) observations. \citet{inn13} further combined these IRSF measurements with OGLE-III $VI$-band photometry to derive the period-Wesenheit relations in various combinations. It is expected that more near infrared data for the SMC Cepheids will be available from the VISTA survey of the Magellanic Clouds System Project \citep[VMC][]{cio11} in the near future.

Using the OGLE-III LMC Cepheid catalog \citep{sos08}, \citet[][hereafter Paper I]{nge09} derived the extinction corrected P-L relations in $VIJHK$ and the four {\it Spitzer} IRAC bands. In this work, we extend our investigation and derive the extinction corrected multi-band P-L relations based on the OGLE-III SMC Cepheids using the catalog from \citet{sos10}. It may be noted that the SMC IRAC bands P-L relations were derived in \citet{nge10} based on single epoch {\it Spitzer} data. In this work the IRAC bands P-L relations are updated using the available photometry up to three epochs. The SMC P-L relation is particularly important in distance scale and stellar pulsation work, such as constraining the theoretical predictions \citep[see][]{bon10}. This is because the metallicity of SMC is $12+\log(O/H)=7.98$~dex, which is similar or comparable to other local dwarf galaxies \citep[such as IC 1613, 7.86~dex; WLM, 7.74~dex; Sextans A, 7.49~dex; Sextans B, 7.56~dex; Pegasus, 7.92~dex; Leo A, 7.38~dex; see][]{sak04,tam11}.

\section{Data and Extinction Correction}

Mean $VI$-band magnitudes and periods for $2626$ fundamental mode SMC Cepheids were taken from \citet{sos10}. The Wesenheit function, $W=I-1.55(V-I)$, was also calculated from the mean $VI$-band magnitudes (if the $V$-band mean magnitude is available). The OGLE-III SMC Cepheids were also matched to the 2MASS point source catalog \citep{cut03,skr06}, using a search radius of $2''$. Mean separation of the $2281$ matched 2MASS sources is $0.225''$, with a dispersion of $0.264''$ (only $60$ matched 2MASS sources have separation greater than $1''$). Random-phase corrections as described in \citet{sos05} were applied to 2MASS photometry to derive the mean $JHK$ magnitudes, using the scaling between $I$-band amplitudes and the $JHK$-band amplitudes. Finally, up to three epochs of the IRAC band photometry, based on publicly released SAGE-SMC \citep[Surveying the Agents of Galaxy Evolution in the Tidally Disrupted, Low-Metallicity Small Magellanic Cloud,][]{gor10} data, were downloaded from the {\it Spitzer} Science Center. As in Paper I and \citet{nge10}, we adopted the SAGE-SMC archival data (version S14 and later, delivered on 2010 September 30) in this work. A search radius of $2''$ was used to match the OGLE-III SMC Cepheids and the sources in SAGE-SMC archival data. The number of matched sources and the corresponding mean separations are summarized in Table \ref{irac_delta} for the SAGE-SMC Epoch 0, 1 and 2 data. Intensity means were calculated using the three epochs data (when available) for each matched Cepheids in the IRAC bands.

As in Paper I, extinction for each OGLE-III SMC Cepheid was estimated using the \citet{zat02} extinction map. For a given input location of SMC Cepheids, this extinction map returns the extinction in $V$-band ($A_V$), measured from the cool stars only. In case the extinction maps did not return any extinction values for a given Cepheid, a mean value of $A_V=0.18$ was adopted. Extinctions in other bands were scaled using the following total-to-selective extinction coefficient: $R_{V,\ I,\ J,\ H,\ K,\ 3.6\mu\mathrm{m},\ 4.5\mu\mathrm{m},\ 5.8\mu\mathrm{m},\ 8.0\mu\mathrm{m}}=\{3.24,\ 1.96,\ 0.95,\ 0.59,\ 0.39,\ 0.17,\ 0.12,\ 0.08,\ 0.05\}$. Following Paper I, the total-to-selective extinction coefficients in $VI$ bands are adopted from \citet{uda99a}, and in other bands these values are calculated based on the extinction law from \citet{car89}. Our values are slightly different from the value of $R_V=3.1$ adopted in \citet{zat02} extinction map \citep{zat99}, and almost identical to those used by \citet{fou07} in $VIJHK$ bands.

\section{The Period-Luminosity Relations}

\begin{figure*}
\plottwo{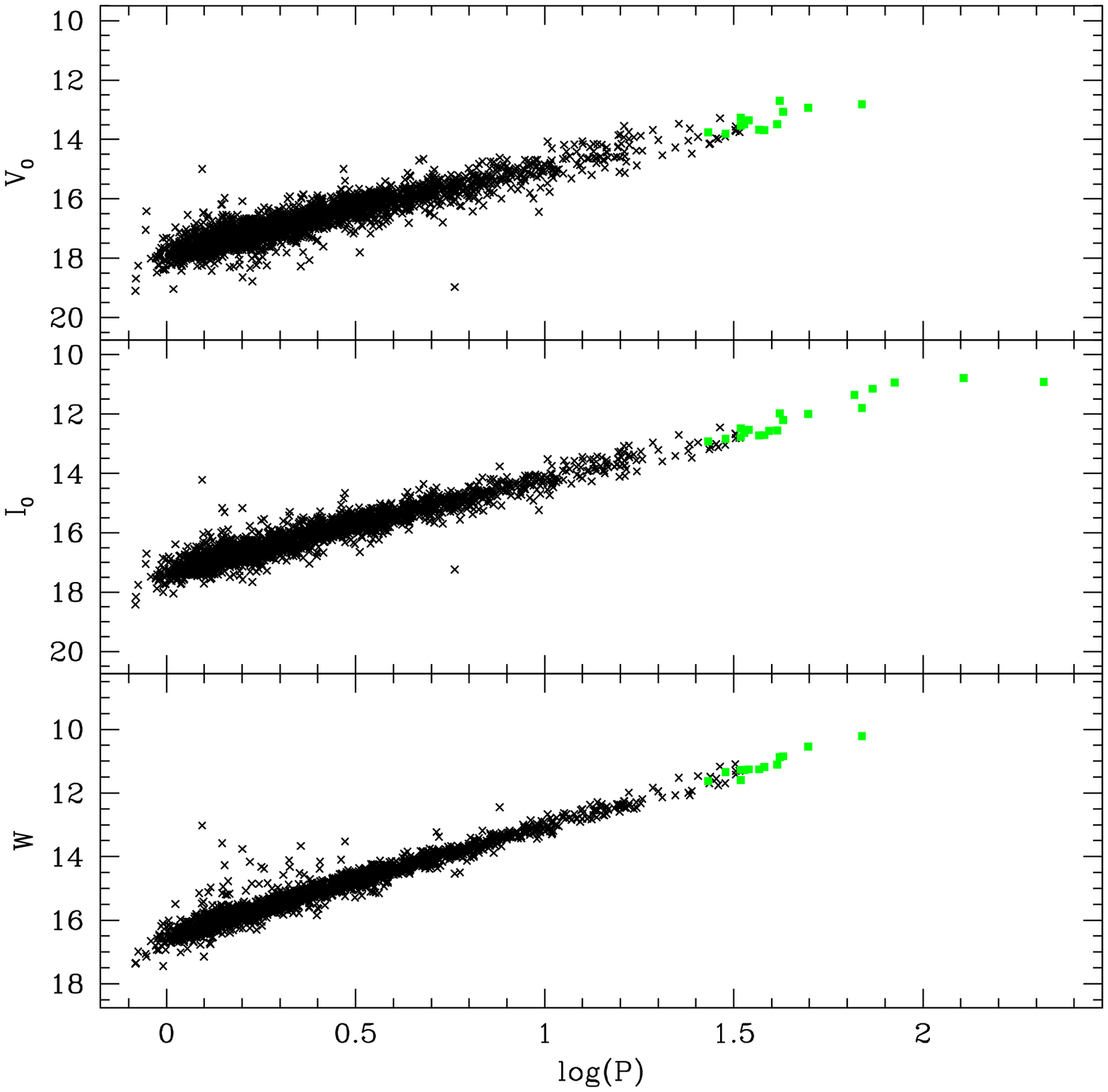}{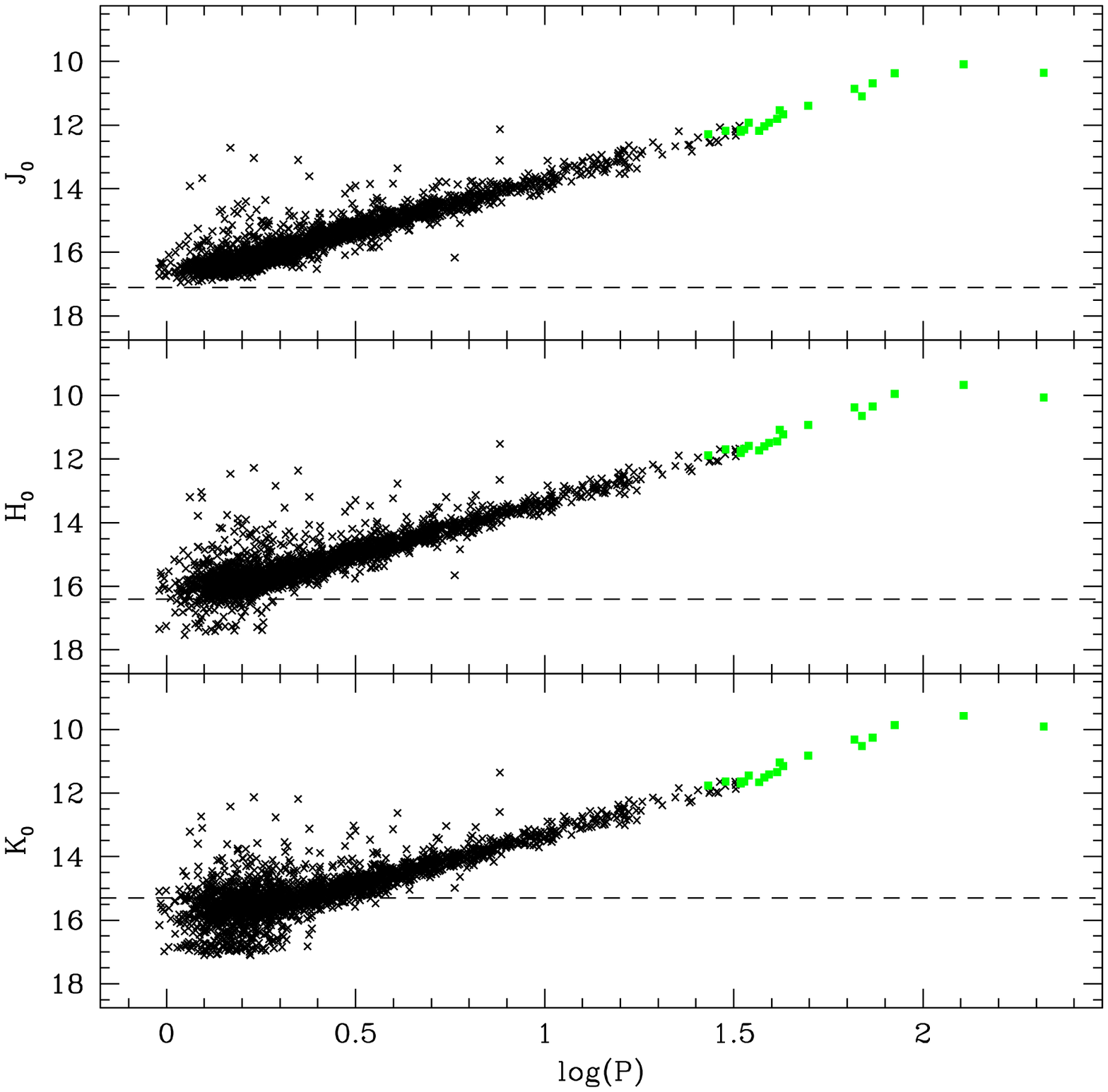}
\caption{The extinction corrected SMC P-L relations in $VI$ bands (left panels) and $JHK$ bands (right panels). The extinction-free Wesenheit function is also included in lower-left panel. The (green) filled squares represent the excluded Cepheids at which the OGLE-III photometry was based on the {\tt DoPHOT} package (see text for more details). The dashed lines in right panels represent the 2MASS $3\sigma$ sensitivity adopted from \citet{cut03}.  \label{pl_smc1}}
\end{figure*} 

\begin{figure}
\epsscale{0.8}
\plotone{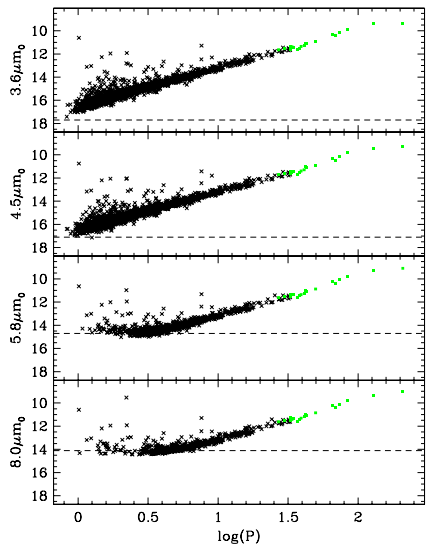}
\caption{The extinction corrected SMC P-L relations in IRAC bands for all the matched sources. The (green) filled squares represent the excluded Cepheids at which the OGLE-III photometry was based on the {\tt DoPHOT} package (see text for more details). The horizontal dashed lines represent the faint limits in each bands, taken from SAGE-SMC document (See {\tt http://data.spitzer.caltech.edu/popular/sage-smc/20100930\newline \_enhanced/documents/sage-smc\_delivery\_sep10.pdf}). 
\label{pl_smc2}}
\end{figure} 

\begin{deluxetable}{lcccc}
\tabletypesize{\scriptsize}
\tablecaption{Summary of the Matched SAGE-SMC Archival Data. \label{irac_delta}}
\tablewidth{0pt}
\tablehead{
\colhead{Band} &
\colhead{$N_{\mathrm{match}}$} &
\colhead{$<\Delta>$\tablenotemark{a}} &
\colhead{Std. Dev.\tablenotemark{b}} &
\colhead{Fraction within $1''$\tablenotemark{c}}
}
\startdata
\cutinhead{Epoch 0} 
$3.6\mu{\mathrm m}$ & 1370 & 0.314 & 0.260 & 97.23\% \\
$4.5\mu{\mathrm m}$ & 1367 & 0.299 & 0.254 & 97.59\% \\
$5.8\mu{\mathrm m}$ &  467 & 0.287 & 0.310 & 95.50\% \\
$8.0\mu{\mathrm m}$ &  361 & 0.281 & 0.317 & 96.12\% \\
\cutinhead{Epoch 1} 
$3.6\mu{\mathrm m}$ & 2580 & 0.272 & 0.241 & 98.22\% \\
$4.5\mu{\mathrm m}$ & 2557 & 0.271 & 0.242 & 98.20\% \\
$5.8\mu{\mathrm m}$ &  702 & 0.276 & 0.299 & 96.30\% \\
$8.0\mu{\mathrm m}$ &  406 & 0.295 & 0.313 & 95.57\% \\
\cutinhead{Epoch 2} 
$3.6\mu{\mathrm m}$ & 2557 & 0.309 & 0.242 & 97.89\% \\
$4.5\mu{\mathrm m}$ & 2545 & 0.308 & 0.238 & 97.92\% \\
$5.8\mu{\mathrm m}$ &  686 & 0.318 & 0.277 & 96.65\% \\
$8.0\mu{\mathrm m}$ &  366 & 0.334 & 0.298 & 95.90\% 
\enddata
\tablenotetext{a}{$\Delta$ is the separation, in arcsecond, between the matched SAGE-SMC Archival sources and the OGLE-III SMC Cepheids.} 
\tablenotetext{b}{The standard deviation of the mean.}
\tablenotetext{c}{Fraction of matched SAGE-SMC Archival sources within $1''$ radius from the OGLE-III SMC Cepheids.}
\end{deluxetable}

Extinction corrected mean magnitudes in each band were used to derive the corresponding P-L relations. The resulting P-L relations are presented in Figures \ref{pl_smc1} and \ref{pl_smc2}. We restricted our Cepheid sample to period range $0.4 < \log(P) < 1.515$ only. \citet{sos10} recommended against using the 17 brightest Cepheids listed in the $I$-band catalog for absolute calibration of brightness. The reason is that these Cepheids were not processed with the standard OGLE Difference Image Analysis (DIA) pipeline as these stars saturate in the reference frames. The shortest period of these 17 Cepheids is $\log(P)\sim1.515$, therefore we adopted an upper limit of the period at $\log(P)=1.515$. Furthermore, Figures \ref{pl_smc1} and \ref{pl_smc2} suggest that the slopes of the P-L relations for Cepheids with $\log(P)>1.9$ become flatter in several bands, as they belong to a sub-class of Cepheids -- the ultra-long period Cepheids \citep{bir09}. The SMC short period Cepheids are known to exhibit a change in the slope of the P-L relation at a specific period. This break period at $\log(P)\sim 0.4$ has been seen in optical data \citep{bau99,uda99a,sha02,san09,sos10,bha2014} and extended to mid-infrared \citep{nge10}. However, \citet{tam11} suggested the break period might occur at $\log(P)=0.55$, while \citet{sub14} found that it could happen at $\log(P)=0.47$. Nevertheless, we adopted $\log(P)=0.4$ as the short period end of our sample. Since our goal is to derive the P-L relations for distance scale applications using the long period Cepheids, detailed investigation into these SMC Cepheids with $\log(P)\leq 0.4$ will be presented elsewhere. Finally, we removed two additional Cepheids (OGLE-SMC-CEP-1157 and OGLE-SMC-CEP-3212), because they were marked in the {\tt remarks.txt} file as their photometry was derived from the {\tt DoPHOT} package rather than the DIA method. 

\begin{figure*}
  $\begin{array}{ccc}
    \includegraphics[angle=0,scale=0.28]{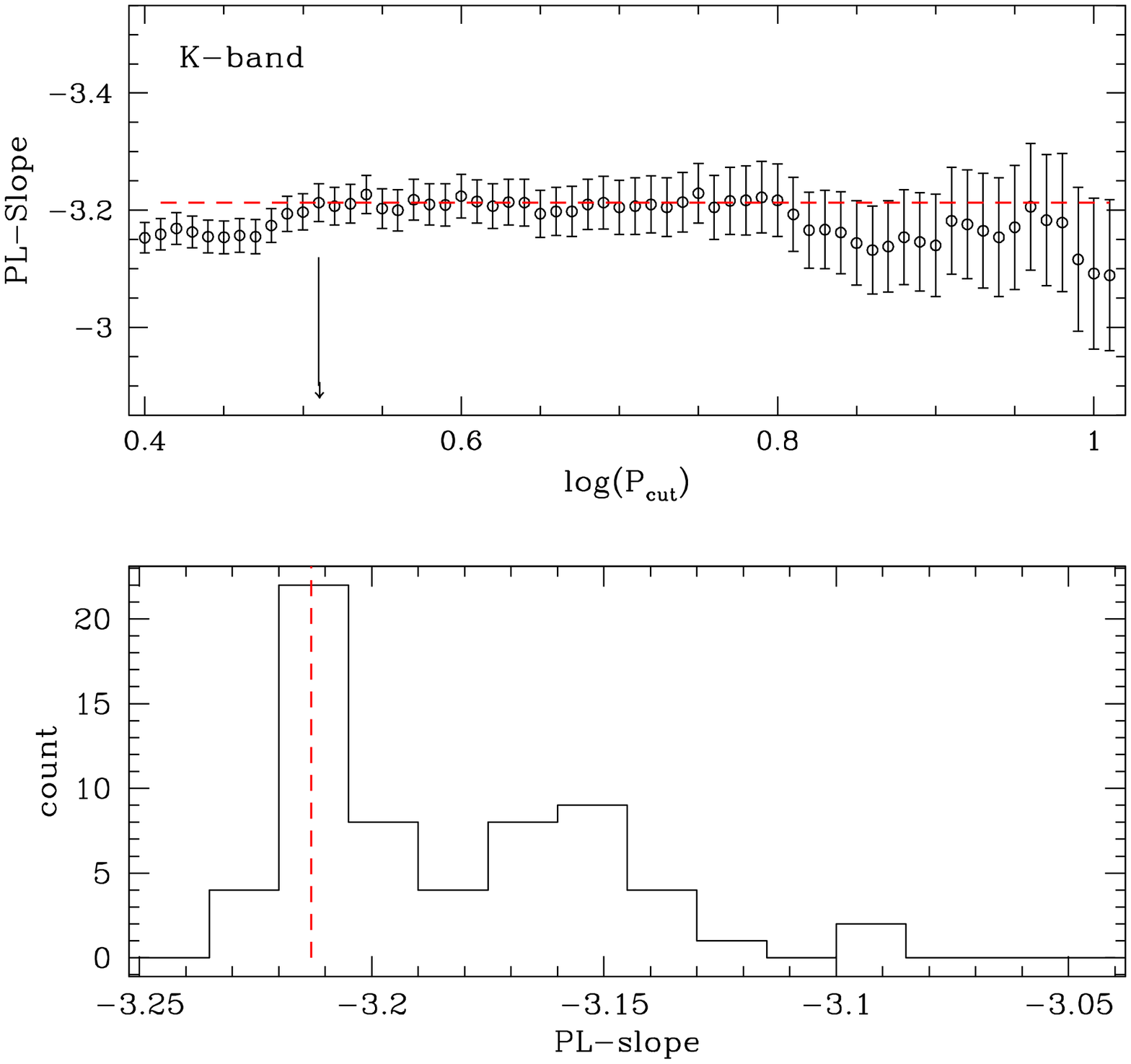} & 
    \includegraphics[angle=0,scale=0.28]{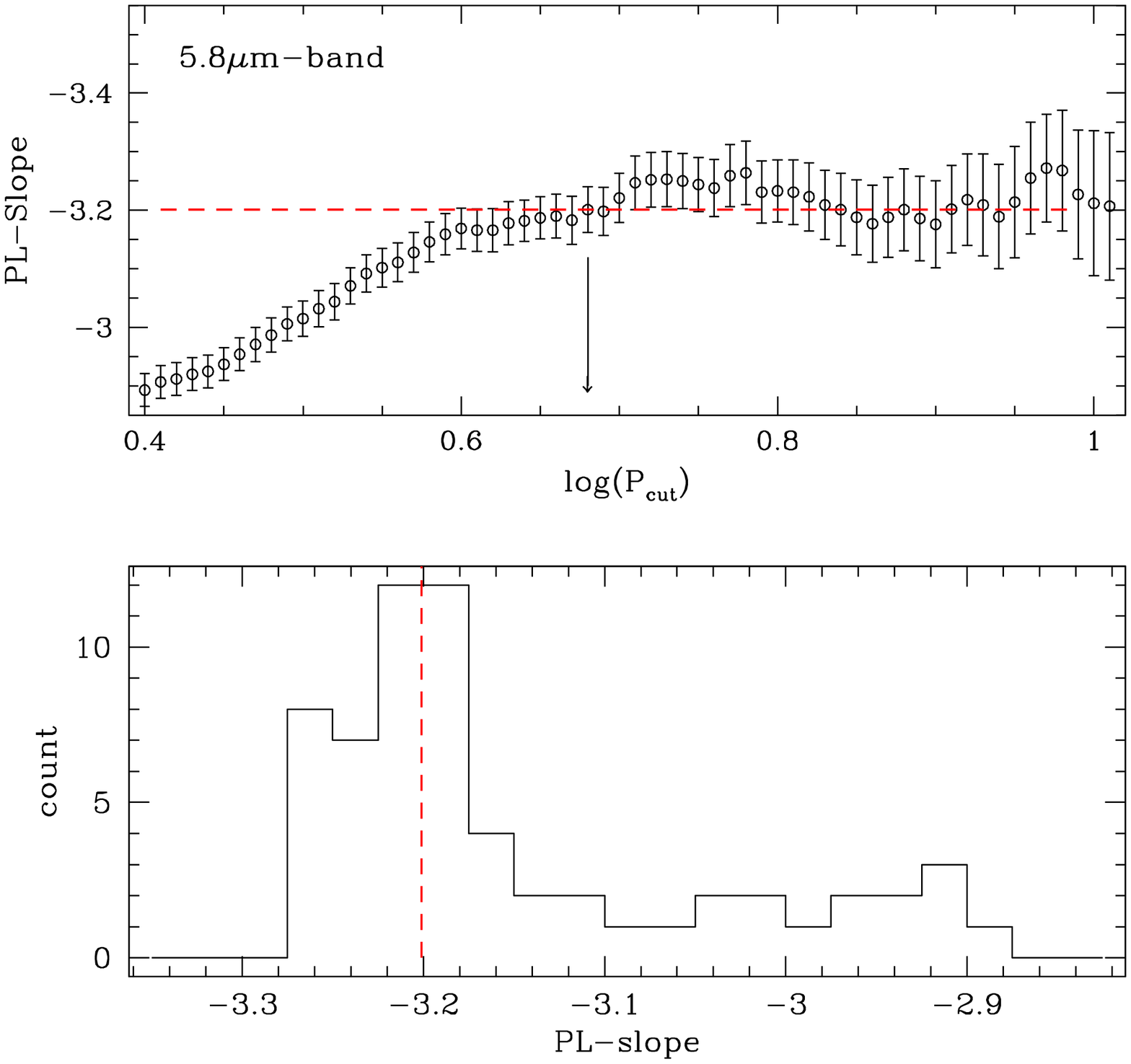} &
    \includegraphics[angle=0,scale=0.28]{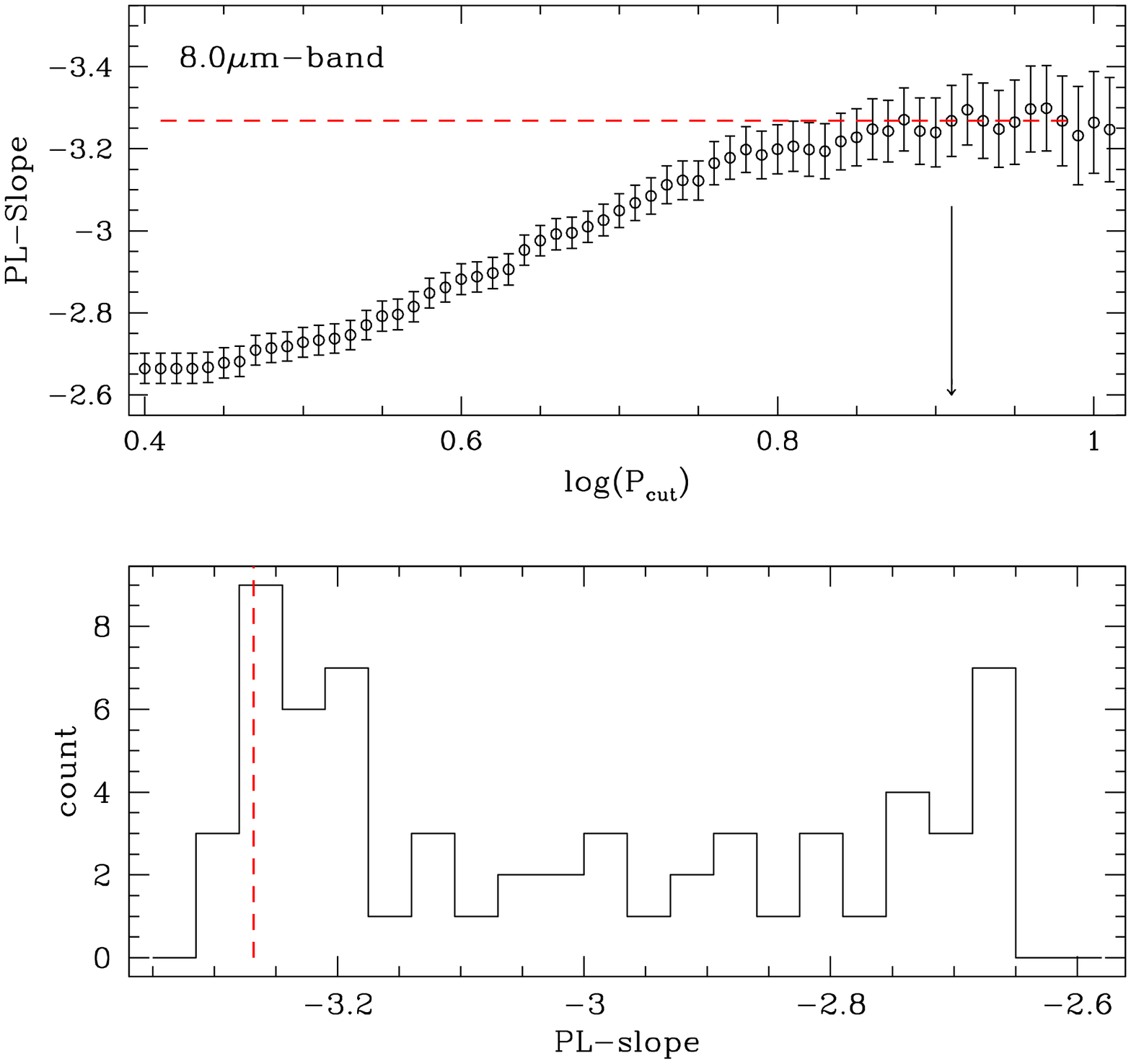} \\ 
  \end{array}$ 
  \caption{The top panels displayed slopes of the fitted P-L relations for data points with $\log(P)>\log(P_{\mathrm{cut}})$ as a function of the adopted period cut. Histograms of the distribution of these fitted P-L slopes were given in the bottom panels. The (red) dashed-lines indicate the mode of the fitted P-L slopes based on the histograms. For $8.0\mu\mathrm{m}$-band P-L slopes, we only derive the mode value from slopes that are smaller than $-3.15$. The vertical arrows represent the adopted $\log(P_{\mathrm{cut}})$ in a given bands. Errors on the P-L slopes are standard errors calculated from the ordinary least squares (OSL) regression method. \label{fig_lpcut}}
\end{figure*}

Inspecting Figures \ref{pl_smc1} and \ref{pl_smc2} suggests that a period cut is needed for $K$-, $5.8$ and $8.0\mu\mathrm{m}$-band P-L relations, after removing $\log(P)\leq 0.4$ Cepheids, to avoid the influence of incompleteness bias at the short-period end. This incompleteness bias is due to the well-known Malmquist bias as discussed, for example, in \citet{san88}. We performed the following steps to determine the appropriate period cuts in the $K$-, $5.8$ and $8.0\mu\mathrm{m}$ bands, respectively.

\begin{enumerate}
\item We first adopted an initial period cut at $\log(P_{\mathrm{cut}})=0.4$ and removed Cepheids having periods smaller than this value.

\item We fitted a P-L relation to the remaining sample of Cepheids with an iterative $2.5\sigma$ clipping algorithm to derive the P-L slopes (and its associated error) for this sample of Cepheids.

\item We repeated step 1 and 2 with a binsize in $\Delta \log P=0.01$, up to a maximum value of $\log(P_{\mathrm{cut}})=1.01$. We picked this binsize such that the parameter space of $\log(P_{\mathrm{cut}})$ can be properly sampled. Upper panels of Figure \ref{fig_lpcut} display the fitted P-L slopes as a function of adopted period cuts in $K$, $5.8$ and $8.0\mu\mathrm{m}$ bands.
  
\item We then plotted the distributions of the P-L slopes given in upper panels of Figure \ref{fig_lpcut} and present these distributions as histograms in the lower panels. 

\item Based on these histograms, we determined the mode of the histogram as an indicator that the P-L slopes begin to stabilize (i.e. without the influence of incompleteness at the short period-end) at a given period cut. The values of the mode are shown as horizontal and vertical dashed (red) line in upper and lower panels of Figure \ref{fig_lpcut}, respectively.

\item We calculated the absolute deviation between the P-L slopes given in the upper panels of Figure \ref{fig_lpcut} and the modal value determined from the histograms (i.e. the absolute deviation between the horizontal dashed line and individual opened circles in the upper panels of Figure \ref{fig_lpcut}).

\item The P-L slope that returned the smallest value of the absolute deviation, based on step 6, is adopted as the final $\log(P_{\mathrm{cut}})$ as indicated by a downward vertical arrow in the upper panels of Figure \ref{fig_lpcut}. In case more than one P-L slope resulted in the smallest value of absolute deviation, we adopted the one with minimum value of $\log(P_{\mathrm{cut}})$.

\end{enumerate}

\noindent Hence, the final adopted $\log(P_{\mathrm{cut}})$ in $K$-, $5.8\mu\mathrm{m}$- and $8.0\mu\mathrm{m}$-band are $0.51$, $0.68$ and $0.91$, respectively. 

\begin{figure}
\plotone{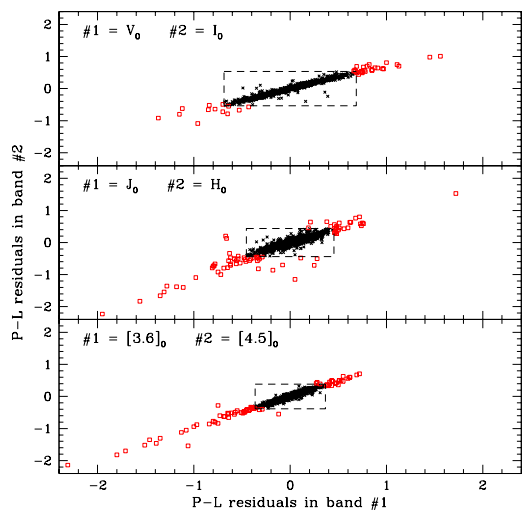}
\caption{Correlations of residuals of P-L relations in two bands. Crosses are the Cepheids remained in the sample after applying the iterative $2.5\sigma$ clipping algorithm, while open (red) squares represent the rejected outliers in the P-L relations (in either bands). The dashed boxes represent the $2.5\sigma$ boundaries to reject the outliers, where $\sigma$ is the dispersion of the P-L relation in a given band. \label{pl_outlier}}
\end{figure}

Following the OGLE team \citep{uda99a,sos10}, obvious outliers of the P-L relations, as displayed in Figures \ref{pl_smc1} and \ref{pl_smc2}, are removed using an iterative $2.5\sigma$ clipping algorithm (cf. Paper I), where $\sigma$ represents the dispersion of the P-L relation in each iteration. These outliers could be caused by a variety of reasons including, but not limited to, blending of nearby sources, mis-matching of the sources and mis-identification as classical Cepheids in the OGLE-III catalog. For example, \citet{nge10} discussed the outliers found in the IRAC bands. Since the goal of this work is to derive the P-L relations, we do not investigate the reasons or nature of these outliers in detail. Nevertheless, it is clear that these outliers should be removed in fitting the P-L relations. Figure \ref{pl_outlier} shows the correlation of residuals of P-L relations in two bands. This demonstrates that the majority of these outliers are present in either or both bands. We have also tried to fit the P-L relations, including the outliers, using the robust regression technique \citep[a regression technique that is robust to outliers presented in the sample,][]{str88,dum89}. The differences obtained in the P-L slopes and zero points using robust regression and our iterative $2.5\sigma$ clipping algorithm do not exceed $0.015$ and $0.012$, respectively, in any given band.

Our final derived P-L relations in various bands for the SMC Cepheids are summarized in Table \ref{tab_pl}. Slopes of the four IRAC bands and the $VIW$-band P-L relations given in Table \ref{tab_pl} are in good agreement with previous determinations as reported in \citet{nge10}, \citet{sos10} and \citet{sub14}. 

\begin{deluxetable}{lccc}
\tabletypesize{\scriptsize}
\tablecaption{Multi-Band P-L Relations for SMC Cepheids. \label{tab_pl}}
\tablewidth{0pt}
\tablehead{
\colhead{Band} & 
\colhead{P-L Slope} &
\colhead{P-L ZP} &
\colhead{P-L Dispersion ($\sigma$)} 
}
\startdata
$V\dots$            & $-2.660\pm0.040$ & $17.606\pm0.028$ & 0.275 \\%& 912  
$I\dots$            & $-2.918\pm0.031$ & $17.127\pm0.022$ & 0.214 \\%& 918 
$J\dots$            & $-3.052\pm0.026$ & $16.773\pm0.019$ & 0.182 \\%& 883 
$H\dots$            & $-3.157\pm0.025$ & $16.530\pm0.018$ & 0.174 \\%& 875 
$K\dots$            & $-3.213\pm0.032$ & $16.514\pm0.025$ & 0.174 \\%& 627  
$3.6\mu{\mathrm m}$ & $-3.220\pm0.021$ & $16.433\pm0.015$ & 0.146 \\%& 881  
$4.5\mu{\mathrm m}$ & $-3.184\pm0.022$ & $16.375\pm0.016$ & 0.155 \\%& 890  
$5.8\mu{\mathrm m}$ & $-3.201\pm0.039$ & $16.362\pm0.036$ & 0.143 \\%& 342  
$8.0\mu{\mathrm m}$ & $-3.268\pm0.087$ & $16.426\pm0.097$ & 0.156 \\%& 134 
$W\dots$            & $-3.314\pm0.020$ & $16.375\pm0.014$ & 0.137 %& 909  
\enddata
\end{deluxetable}

\subsection{Comparison with P-L Relations from OGLE-II}

\begin{deluxetable}{lccccc}
\tabletypesize{\scriptsize}
\tablecaption{Comparison of SMC P-L Relations. \label{tab_compare}}
\tablewidth{0pt}
\tablehead{
\colhead{Ref.} & 
\colhead{P-L Slope} &
\colhead{P-L ZP} &
\colhead{$N$} &
\colhead{$|T|$} &
\colhead{$p$-value} 
}
\startdata
\multicolumn{6}{c}{$V$-band}\\
1            & $-2.660\pm0.040$ & $17.606\pm0.028$ & 912 & $\cdots$ & $\cdots$ \\
2            & $-2.572\pm0.042$ & $17.480\pm0.032$ & 466 & 1.495 & 0.135 \\
3            & $-2.573\pm0.041$ & $17.492\pm0.032$ & 464 & 1.492 & 0.136 \\
4            & $-2.588\pm0.045$ & $17.530\pm0.035$\tablenotemark{a} & 460 & 1.191 & 0.234 \\
5            & $-2.590\pm0.047$ & $17.600\pm0.019$ & 488 & 1.145 & 0.253 \\
\multicolumn{6}{c}{$I$-band}\\
1            & $-2.918\pm0.031$ & $17.127\pm0.022$ & 918 & $\cdots$ & $\cdots$ \\
2            & $-2.857\pm0.033$ & $17.039\pm0.025$ & 488 & 1.336 & 0.182 \\
3            & $-2.843\pm0.033$ & $17.052\pm0.025$ & 487 & 1.638 & 0.102 \\
4            & $-2.862\pm0.035$ & $17.083\pm0.027$\tablenotemark{a} & 462 & 1.192 & 0.233 \\
5            & $-2.865\pm0.036$ & $17.117\pm0.014$ & 488 & 1.122 & 0.262 \\
\multicolumn{6}{c}{$J$-band}\\
1            & $-3.052\pm0.026$ & $16.773\pm0.019$ & 883 & $\cdots$ & $\cdots$ \\
6            & $-3.037\pm0.034$ & $16.771\pm0.027$ & 418 & 0.359 & 0.720 \\
\multicolumn{6}{c}{$H$-band}\\
1            & $-3.157\pm0.025$ & $16.530\pm0.018$ & 875 & $\cdots$ & $\cdots$ \\
6            & $-3.160\pm0.032$ & $16.475\pm0.025$ & 414 & 0.075 & 0.940 \\
\multicolumn{6}{c}{$K$-band}\\
1            & $-3.213\pm0.032$ & $16.514\pm0.025$ & 627 & $\cdots$ & $\cdots$ \\
6            & $-3.212\pm0.033$ & $16.494\pm0.026$ & 418 & 0.022 & 0.983  \\
\multicolumn{6}{c}{$W$-band}\\
1            & $-3.314\pm0.020$ & $16.375\pm0.014$ & 909 & $\cdots$ & $\cdots$ \\
2            & $-3.303\pm0.022$ & $16.345\pm0.017$ & 469 & 0.369  & 0.712  \\
3            & $-3.310\pm0.020$ & $16.387\pm0.016$ & 463 & 0.139  & 0.890  \\
7            & $-3.300\pm0.021$ & $16.381\pm0.016$ & 446 & 0.473  & 0.636  
\enddata
\tablecomments{{\bf Reference:} (1). this work; (2) \citet{uda99a}; (3). same as (2) but updated in {\tt ftp://sirius.astrouw.edu.pl/ogle/\\ ogle2/var\_stars/smc/cep/catalog/README.PL}; (4) \citet{san09}; (5) \citet{sto04}; (6) \citet{gro00}; (7) \citet{gro00}, with $2.5\sigma$-clipping.}
\tablenotetext{a}{By adopting $\mu_{SMC}=18.93$~mag.}
\end{deluxetable}

The multi-band P-L relations based on the OGLE-III SMC Cepheids are compared to the P-L relations derived from the OGLE-II catalog in Table \ref{tab_compare}. As in Paper I, we applied the $t$-test to test the consistency of P-L slopes derived here to the published values. The calculated $T$-values and the corresponding $p$-values (the probability), evaluated based on the $t$-distribution with $\alpha=0.05$ (where $\alpha$ is the adopted significance level), are listed in the last two columns of Table \ref{tab_compare}. The expected $t$-value from the $t$-distribution, which only depends on $\alpha$ and the degrees of freedom ($\nu$), is $1.968$ for $\nu=300$ or $1.962$ for $\nu=1400$ (which roughly covers the range of $\nu$ in our test). The null hypothesis is that the P-L slopes given in Table \ref{tab_pl} are the same as the P-L slopes based on OGLE-II Cepheids. This can be rejected if $|T|>t$, where $t\sim1.96$ (or equivalently, $p$-value smaller than $0.05$). From Table \ref{tab_compare}, the P-L slopes in all bands are consistent with those derived from OGLE-II catalogs. We did not consider the updated SMC P-L slopes given in \citet{uda00} as the author assumed the P-L slopes are the same in both LMC and SMC. For $VI$-band P-L relations, we note that different photometric reduction packages were used in OGLE-II \citep[{\tt DoPHOT}, see][]{uda98,uda00} and OGLE-III \citep[DIA, see][]{uda08}. For common Cepheids with $\log (P)>0.4$ in OGLE-II and OGLE-III catalogs, the averaged difference in mean magnitudes (i.e. OGLE-II minus OGLE-III) is $\sim0.009$~mag and $\sim0.004$~mag in $V$- and $I$-band, respectively. After including extinction corrections, the averaged differences increased to $\sim -0.093$~mag in the $V$-band and $\sim -0.057$~mag in the $I$-band. In addition, these differences do not show any dependency on the pulsation period. 

\section{Testing for Nonlinear P-L Relations at 10 Days}

\begin{deluxetable*}{lcccccccc}
\tabletypesize{\scriptsize}
\tablecaption{SMC P-L Relations Separated at 10~Days. \label{tab_ftest}}
\tablewidth{0pt}
\tablehead{
\colhead{Band} & 
\colhead{P-L Slope$_S$} &
\colhead{P-L ZP$_S$} &
\colhead{$\sigma_S$} &
\colhead{$N_S$} & 
\colhead{P-L Slope$_L$} &
\colhead{P-L ZP$_L$} &
\colhead{$\sigma_L$} &
\colhead{$N_L$} 
}
\startdata
$V\dots$            & $-2.634\pm0.061$ & $17.592\pm0.039$ & 0.269 & 821 & $-2.453\pm0.244$ & $17.345\pm0.290$ & 0.328 & 91 \\ 
$I\dots$            & $-2.888\pm0.048$ & $17.127\pm0.022$ & 0.209 & 825 & $-2.808\pm0.185$ & $16.981\pm0.220$ & 0.248 & 93 \\ 
$J\dots$            & $-3.008\pm0.041$ & $16.748\pm0.026$ & 0.180 & 790 & $-2.991\pm0.144$ & $16.684\pm0.171$ & 0.195 & 93 \\ 
$H\dots$            & $-3.114\pm0.040$ & $16.506\pm0.025$ & 0.173 & 781 & $-3.149\pm0.131$ & $16.505\pm0.156$ & 0.178 & 94 \\ 
$K\dots$            & $-3.203\pm0.057$ & $16.508\pm0.040$ & 0.173 & 532 & $-3.092\pm0.129$ & $16.364\pm0.153$ & 0.176 & 95 \\ 
$3.6\mu{\mathrm m}$ & $-3.185\pm0.034$ & $16.413\pm0.021$ & 0.146 & 789 & $-3.259\pm0.112$ & $16.468\pm0.134$ & 0.152 & 92 \\ 
$4.5\mu{\mathrm m}$ & $-3.166\pm0.035$ & $16.365\pm0.023$ & 0.155 & 797 & $-3.108\pm0.115$ & $16.276\pm0.137$ & 0.156 & 93 \\ 
$5.8\mu{\mathrm m}$ & $-3.153\pm0.098$ & $16.324\pm0.080$ & 0.143 & 252 & $-3.229\pm0.109$ & $16.394\pm0.130$ & 0.145 & 90 \\ 
$8.0\mu{\mathrm m}$ & $-2.946\pm0.909$ & $16.121\pm0.872$ & 0.146 &  42 & $-3.254\pm0.121$ & $16.409\pm0.143$ & 0.161 & 92 \\ 
$W\dots$            & $-3.320\pm0.031$ & $16.378\pm0.019$ & 0.136 & 817 & $-3.274\pm0.115$ & $16.329\pm0.136$ & 0.151 & 92 
\enddata
\tablecomments{The subscripts $_S$ and $_L$ stand for Cepheids with $0.4<\log(P)<1.0$ (i.e. short period Cepheids) and $\log(P)>1.0$ (i.e. long period Cepheids), respectively. $ZP$ and $\sigma$ represents the zero point and dispersion of the P-L relation, respectively. $N$ is the number of Cepheids used in deriving the P-L relations.}
\end{deluxetable*}

The LMC P-L relation is known to exhibit a break at a period around 10~days, as shown in Paper I (and references therein) and \citet{bha2015} with rigorous statistical tests. \citet{san09} proposed that the same break should occur for the SMC P-L relations in optical bands based on an analysis of the period-color relations. However their derived SMC P-L relations do not exhibit a break at 10~days. In the $JHK$-bands, \citet{mat11} did not find any significant break at 10~days using the single-epoch data taken from the IRSF. In contrast, studies done in \citet{bon10} and \citet{gar13} suggested that the SMC P-L relations should be nonlinear. In this section we examine the nonlinearity of the SMC P-L relations based on our data presented in previous section, where nonlinearity refers to an underlying P-L relation can be separated to two P-L relations with a break period at 10~days. Table \ref{tab_ftest} presents the fitted SMC P-L relations for long and short period Cepheids separated at 10~days.

\begin{figure*}
  $\begin{array}{ccc}
    \includegraphics[angle=0,scale=0.28]{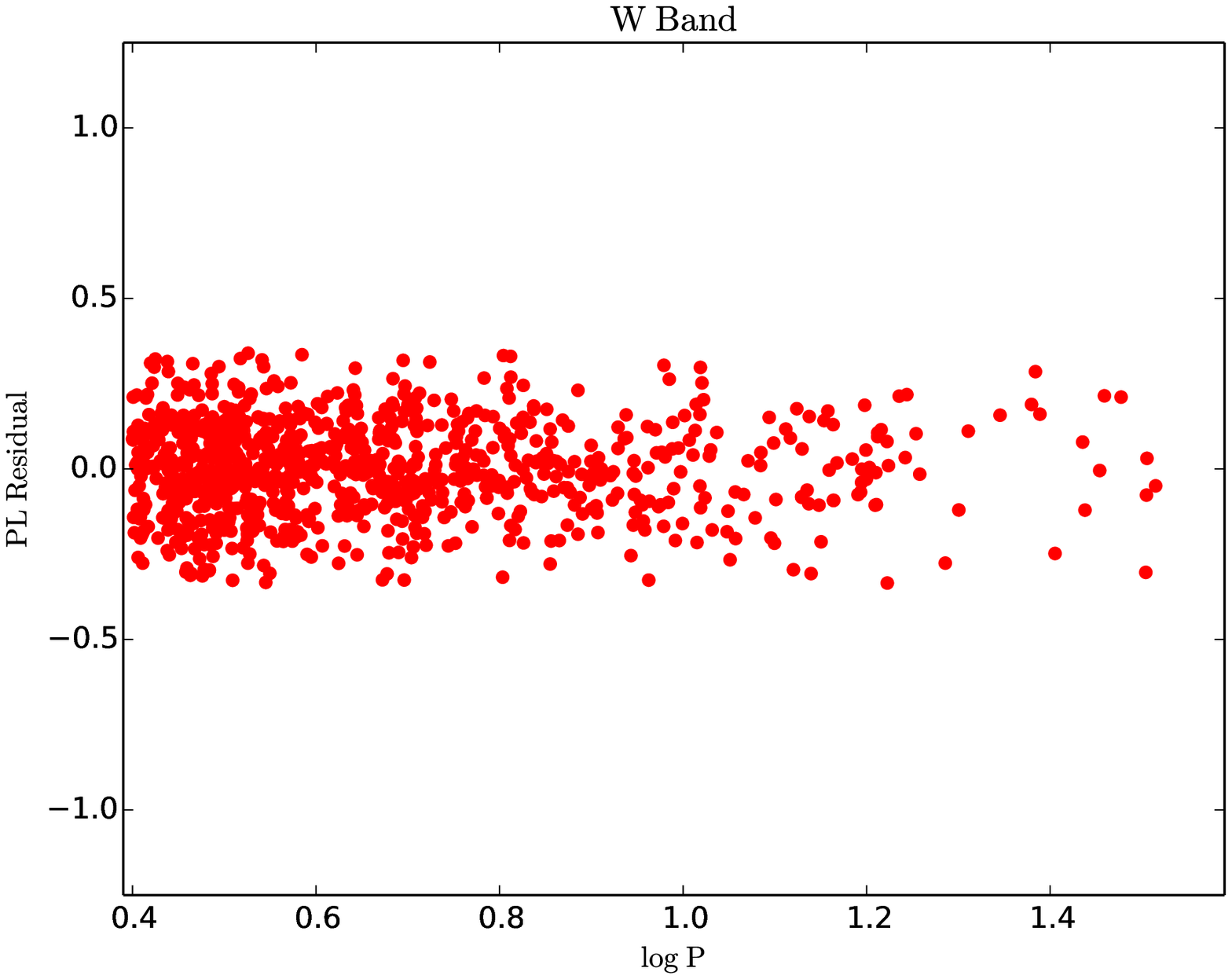} &
    \includegraphics[angle=0,scale=0.28]{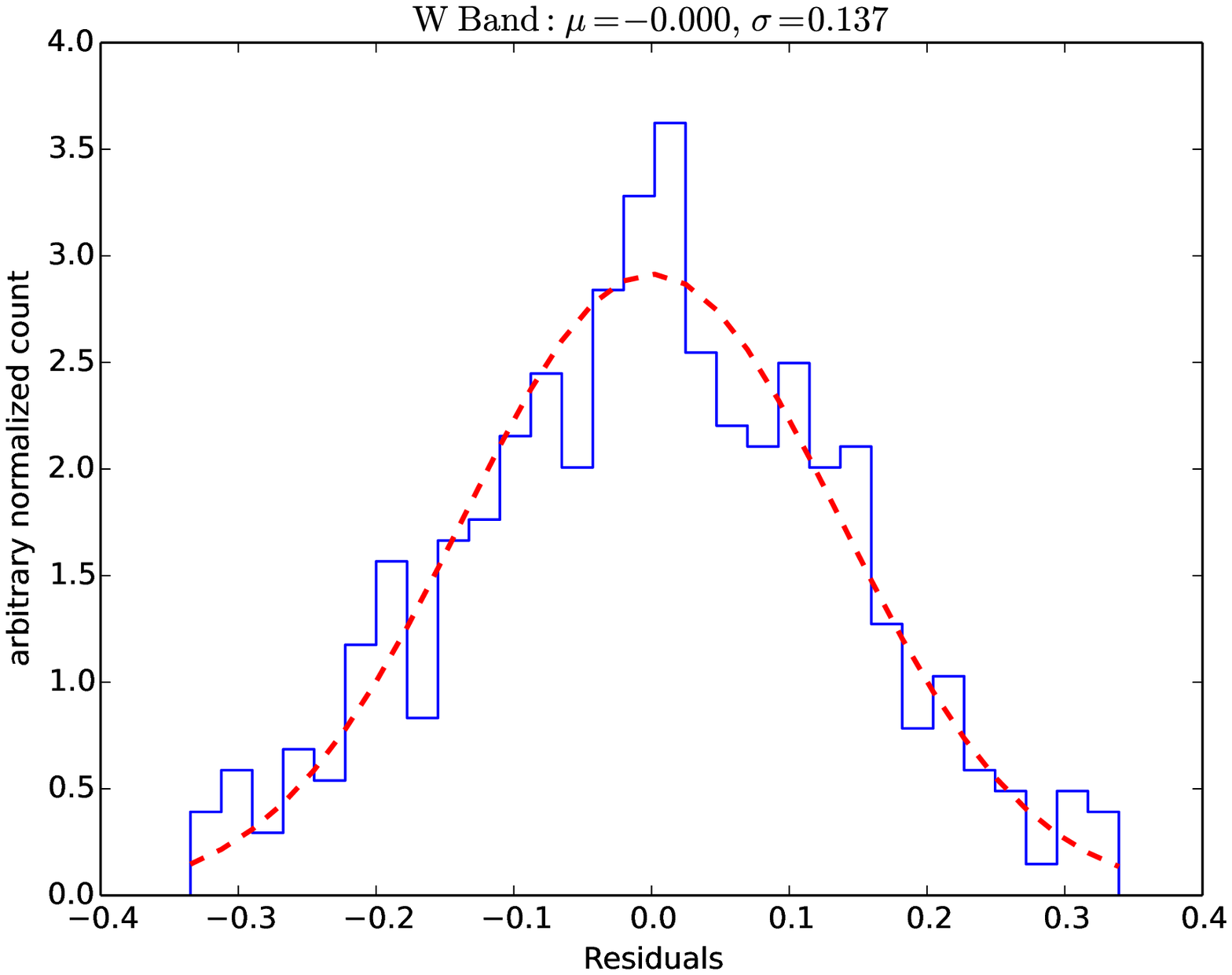} & 
    \includegraphics[angle=0,scale=0.28]{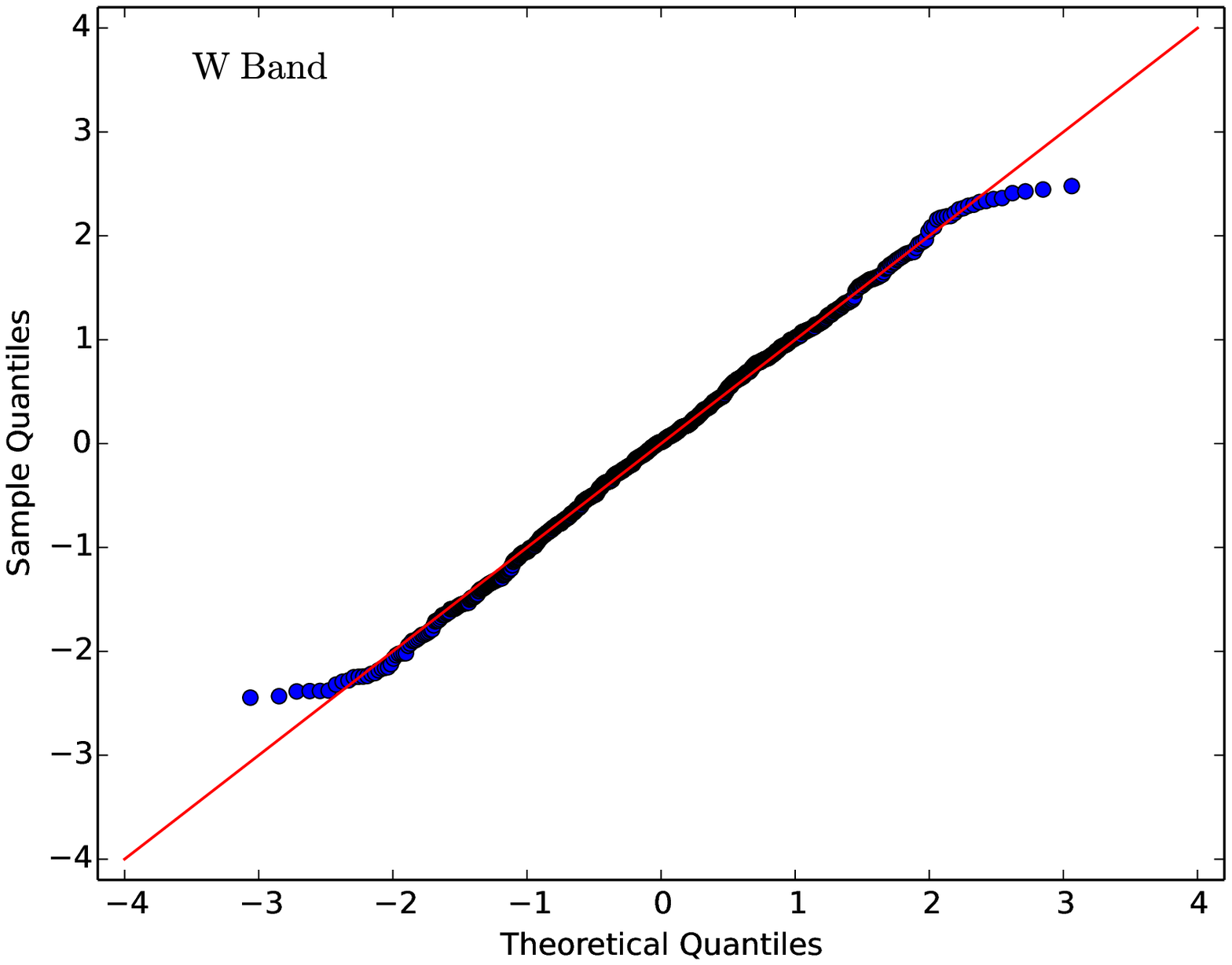} \\    
  \end{array}$ 
  \caption{{\it Left Panel:} residuals of the fitted P-L relation in Wesenheit function as a function of logarithmic period. {\it Middle Panel:} distribution of the residuals presented in the left panel. The fitted Gaussian function to the histogram is shown as dashed (red) curve, with fitted Gaussian parameters $(\mu,\ \sigma)$ given at top of the plot. {\it Right Panel:} the $qq$ plot of the residuals. The (red) line represents the case of $y=x$. The theoretical quantiles represent the quantiles if the residuals are drawn from a Gaussian distribution, while the sample quantiles are actual quantiles calculated from the residuals. } \label{fig_res}
\end{figure*}

As pointed out by \citet{nge06}, statistical tests are needed to find out the nonlinearity of P-L relation. We apply the $F$-test (as in Paper I) and a random walk method in this work to examine the nonlinearity of multi-band SMC P-L relations with a break period at 10~days. The $F$-test requires independent and identically distributed (iid) variables, as well as homoscedasticity and normality of residuals. Since observations of one Cepheid are independent of observations of other Cepheids, the iid assumption is satisfied. We next examine the homoscedasticity assumption (i.e. constant variance) of the P-L residuals. The left panel of Figure \ref{fig_res} displays the residuals of P-L relation: these do not exhibit any obvious trends that violate the homoscedasticity assumption. Further, the average of the P-L residuals are found to be consistent with zero in all bands. Finally, we tested the assumption of normality of the P-L residuals. The middle and right panels of Figure \ref{fig_res} show the histogram and the quantile-quantile ($qq$) plot of the P-L residuals, respectively. The distribution of residuals can be represented with a Gaussian function, as indicated by a dashed curve in middle panel of Figure \ref{fig_res}. The $qq$ plot is a common diagnostic tool to evaluate the normality assumption: the quantiles of the data should fall in a diagonal straight line (i.e. the $y=x$ line). Our $qq$ plot demonstrates that the majority of the P-L residuals follow a normal distribution (except few points at the extreme ends of $qq$ plot). Even though in Figure \ref{fig_res} we only showed the results of P-L residuals from the Wesenheit function, P-L residuals in other bands all look very similar to this Figure. Normality of the residuals is important in assuring that the distribution of the $F$-test statistic under the null hypothesis of linearity follows the $F$ distribution. However we can also use a permutation method \citep[as describe below,][]{kim00} to generate the distribution of the $F$ statistic in a way that is independent of the distribution of the residuals. 

We fit a single regression line to the actual data and then fit two lines with a break point at a period of 10~days, followed by calculating the $F$-test statistic, $F_0$. Details of the $F$-test can be found in Paper I and reference therein, and will not be repeated here. In short, a $F_0$ value that is greater than $\sim3$ would indicate that the underlying P-L relation is nonlinear. If we have $N$ data points we can calculate $N$ residuals from the single straight line fit, ${\epsilon}_i$ (where $i=1,\cdots,N$). Next we randomly permute the residuals ${\epsilon}_i = {\epsilon}_j$ and add the permuted residuals back to the fitted data from the single straight line fit. This generates one iteration of pseudo-data. Note the $i^{th}$ data point still has the original period. Now with this pseudo-data, we fit a straight line and then two lines with the original break point and calculate the $F$-test statistic, $F_1$. We repeat this process 10,000 times to obtain 10,000 independent $F$-test statistics. We then find the proportion of the $F_i$ that are greater than the observed $F_0$. This proportion gives the $p$-value, or significance, of the observed $F_0$-test statistic without assuming normality. Results of the $F$-test statistic, $F_0$ and the $p$-value calculated using the permutation method are summarized in Table \ref{tab_new} for the SMC P-L relations. The largest $F_0$-value for the SMC P-L relations is found to be $1.17$ in the $J$-band. Therefore, our $F$-test results strongly suggest that the SMC P-L relations are linear from optical to infrared bands. Our results are further confirmed with a non-parametric random walk test \citep[see][]{koe2007,bha2015}. A $p(R)$-value from the random walk test such that $p(R) \gtrsim 0.1$ would indicate that the null hypothesis of cannot be rejected. We found that $p(R)>0.3$ in all cases with the exception in the $H$-band P-L relation, which displays a marginal $p$-value (0.06) from the random walk test. We also emphasize that the random walk test is non-parametric and does not make any assumptions about homoscedasticity or normality of residuals. Since the short period Cepheids in $8.0\mu\mathrm{m}$-band only occupied a small period range, $0.91 < \log (P) < 1.0$, that will cause the determination of the P-L slope to be less accurate ($-2.946\pm0.909$, see Table \ref{tab_ftest}), we excluded the $8.0\mu\mathrm{m}$-band P-L relation from our statistical tests.

\begin{deluxetable}{lcc}
\tabletypesize{\scriptsize}
\tablecaption{$F$-test and Permutation Method Results for the Nonlinearity of P-L Relations. \label{tab_new}}
\tablewidth{0pt}
\tablehead{
\colhead{Band} & 
\colhead{$F_0$} &
\colhead{$p$} 
}
\startdata
$V\dots$            & 0.786 & 0.454 \\
$I\dots$            & 0.681 & 0.504 \\
$J\dots$            & 1.165 & 0.315 \\
$H\dots$            & 0.959 & 0.395 \\
$K\dots$            & 0.559 & 0.580 \\
$3.6\mu{\mathrm m}$ & 0.888 & 0.414 \\
$4.5\mu{\mathrm m}$ & 0.526 & 0.597 \\
$5.8\mu{\mathrm m}$ & 0.162 & 0.846 \\
$W\dots$            & 0.095 & 0.908
\enddata
\end{deluxetable}

\section{Comparison with LMC P-L Relations and the Universal Period-Wesenheit Relation}

\begin{deluxetable}{lccccc}
\tabletypesize{\scriptsize}
\tablecaption{Comparison of the P-L slopes Between LMC and SMC Cepheids. \label{tab_lmcsmc}}
\tablewidth{0pt}
\tablehead{
\colhead{Band} & 
\colhead{Galaxy} & 
\colhead{P-L Slope} &
\colhead{$|\Delta|$\tablenotemark{a}} & 
\colhead{$|T|$} &
\colhead{$p$-value} 
}
\startdata

$V\dots$    & SMC            & $-2.660\pm0.040$ & \multirow{2}{*}{$0.109\pm0.046$} & \multirow{2}{*}{2.558} & \multirow{2}{*}{0.011} \\
            & LMC            & $-2.769\pm0.023$ &  \\ % 2.37sigma, T>t
$I\dots$    & SMC            & $-2.918\pm0.031$ & \multirow{2}{*}{$0.043\pm0.034$} & \multirow{2}{*}{1.407} & \multirow{2}{*}{0.160} \\
            & LMC            & $-2.961\pm0.015$ &  \\ % 1.26sigma, T>t
$J\dots$    & SMC            & $-3.052\pm0.026$ & \multirow{2}{*}{$0.063\pm0.030$} & \multirow{2}{*}{2.351} & \multirow{2}{*}{0.019} \\
            & LMC            & $-3.115\pm0.014$ &  \\ % 2.1sigma, T>t
$H\dots$    & SMC            & $-3.157\pm0.025$ & \multirow{2}{*}{$0.049\pm0.028$} & \multirow{2}{*}{1.930} & \multirow{2}{*}{0.054} \\
            & LMC            & $-3.206\pm0.013$ &  \\ % 1.75sigma, T>t
$K\dots$    & SMC            & $-3.213\pm0.032$ & \multirow{2}{*}{$0.019\pm0.035$} & \multirow{2}{*}{0.616} & \multirow{2}{*}{0.538} \\
            & LMC            & $-3.194\pm0.015$ &  \\ % 0.54sigma, T>t
$3.6\mu\mathrm{m}$ & SMC            & $-3.220\pm0.021$ & \multirow{2}{*}{$0.033\pm0.023$} & \multirow{2}{*}{1.609} & \multirow{2}{*}{0.108} \\
                   & LMC            & $-3.253\pm0.010$ &  \\ % 1.43sigma, T>t
$4.5\mu\mathrm{m}$ & SMC            & $-3.184\pm0.022$ & \multirow{2}{*}{$0.030\pm0.024$} & \multirow{2}{*}{1.421} & \multirow{2}{*}{0.155} \\
                   & LMC            & $-3.214\pm0.010$ &  \\ % 1.25sigma, T>t
$5.8\mu\mathrm{m}$ & SMC            & $-3.201\pm0.039$ & \multirow{2}{*}{$0.019\pm0.044$} & \multirow{2}{*}{0.467} & \multirow{2}{*}{0.641} \\
                   & LMC            & $-3.182\pm0.020$ &  \\ % 0.43sigma, T<t
$8.0\mu\mathrm{m}$ & SMC            & $-3.268\pm0.087$ & \multirow{2}{*}{$0.071\pm0.094$} & \multirow{2}{*}{0.805} & \multirow{2}{*}{0.421} \\
                   & LMC            & $-3.197\pm0.036$ &   \\ % 0.76sigma, T>t
$W\dots$   & SMC            & $-3.314\pm0.020$ & \multirow{2}{*}{$0.001\pm0.022$} & \multirow{2}{*}{0.055} & \multirow{2}{*}{0.957} \\
           & LMC            & $-3.313\pm0.008$ &    % 0.05sigma, T<t
\enddata
\tablenotetext{a}{$\Delta$ is the difference of the slopes between LMC and SMC P-L relations. The error for $\Delta$, denoted as $\sigma_S$, is the quadrature sum of the errors in two slopes.}
\end{deluxetable}

Since the data sources, algorithms and codes used in this work are almost identical to Paper I, we can compare the P-L slopes found in this work for the SMC Cepheids to the \emph{linear} version of the LMC P-L slopes given in Paper I. This provides a critical test for the assumption of a universal P-L slopes across different filters. As in the previous section, we applied the statistical $t$-test to test the null hypothesis such that P-L slopes from LMC and SMC Cepheids be the same in a given band. Table \ref{tab_lmcsmc} summarizes our $t$-test results. We also calculated the (absolute) difference of these P-L slopes, $|\Delta|$, and its associated error ($\sigma_S$), which are also provided in Table \ref{tab_lmcsmc}. When comparing the two slopes, it is a common practice in the literature to claim the two underlying slopes as consistent if $\Delta$ is within, say, $\sim2.5\sigma_S$. In this case, the LMC and SMC P-L slopes listed in Table \ref{tab_lmcsmc} are consistent with each other in all bands. Based on the $t$-test results presented in Table \ref{tab_lmcsmc}, the null hypothesis of equivalent slopes can be rejected for the P-L slopes in $V$- and $J$-band only. The $H$-band P-L slopes, on the other hand, provide evidence of marginal consistency of the slopes. Therefore, the slopes of the P-L relations are not universal in $VJ$ bands, at least for metallicity bracketed by the Magellanic Clouds.  

On the other hand, the Wesenheit function, that incorporates a color term, provides evidence of a remarkably consistent P-L slope between the LMC and SMC Cepheids with a value of $\sim -3.31$. This strongly indicates that the Wesenheit function is universal \citep[for example, see discussion in][]{bon10}, and hence more suitable for the distance scale application than the single band P-L relations. Since the P-L slopes for Wesenheit function are almost identical, the difference of the P-L zero points for LMC and SMC Wesenheit functions directly translates to the relative distance between LMC and SMC. In terms of distance modulus $\mu$, it is found to be $\Delta \mu= 0.483 \pm 0.015$~mag, which is in good agreement with the value found in \citet[][$\Delta \mu= 0.48\pm0.03$~mag based on fundamental mode Cepheids only]{inn13} or the preferred value based on Gaussian fitting to a number of recent measurements \citep[][$\Delta \mu= 0.458 \pm 0.068$~mag]{gra14}. The recommended distance moduli for LMC and SMC are $18.49\pm0.09$~mag \citep{deg14} and $18.96\pm0.02$~mag \citep{deg15}, respectively, with a difference of $\Delta \mu = 0.47\pm0.09$~mag. Again this value is consistent with our result. By shifting the SMC data with $0.483$~mag, we combine the Cepheids in both Magellanic Clouds and derive the following P-L relation for $2578$ Cepheids:

\begin{eqnarray}
  W & = & -3.314 (\pm0.009) \log P + 15.892 (\pm 0.006), \ \ \sigma=0.099, \nonumber
\end{eqnarray}

\noindent which, as expected, is identical to the LMC period-Wesenheit relation given in Paper I.

\section{Conclusion}

The main goal of this study was to extend the work of Paper I by deriving the multi-band P-L relations for SMC fundamental mode Cepheids using the latest compilation of OGLE-III catalog \citep{sos10}. In addition to the $VI$-band mean magnitudes adopted from \citet{sos10}, we also matched the OGLE-III SMC Cepheids to 2MASS and SAGE-SMC catalogs and derived the mean magnitudes in infrared $JHK$-band and the four IRAC bands. Extinction corrections to these mean magnitudes were done using the \citet{zat02} extinction map. These data sources are the same, or very similar, to those adopted in Paper I for the LMC Cepheids. We also applied the almost identical algorithms and codes from Paper I in this work. The difference between this work and Paper I, on the contrary, includes (a) Cepheids with $\log(P)<0.4$ and some longest period Cepheids were excluded in the SMC sample; and (b) we applied a period cut to the $K$, $5.8\mu\mathrm{m}$- and $8.0\mu\mathrm{m}$-band P-L relations to avoid the bias due to incompleteness at the faint end.

We then derived the extinction corrected P-L relations in $VIJHK$ bands and in the four IRAC bands, as well as the extinction free period-Wesenheit function, for SMC fundamental mode Cepheids. Following Paper I, we did not apply absolute calibration to our P-L relation because the readers can adopt their preferred SMC distance to calibrate these P-L relations \citep[for example, the latest measurement of SMC distance can be found in][]{gra14}. We summarize the main results based on the derived SMC P-L relations as follows:

\begin{enumerate}
\item Based on the $F$-test, the SMC P-L relations are found to be linear from optical to infrared bands (except the $8.0\mu\mathrm{m}$-band P-L relation at which the $F$-test is not applied) for SMC Cepheids with period between $\log (P)=0.40$ to $\log (P)=1.51$ (for $K$-band, the period range is $0.51<\log (P)<1.51$; for $5.8\mu\mathrm{m}$ band, the period range is $0.68<\log (P)<1.51$). The period-Wesenheit relation is also linear in the same period range. 
\item Based on the $t$-test, the null hypothesis of equivalent slopes for the LMC and SMC P-L relations can be rejected in the $VJ$ bands. The P-L slopes in other bands are consistent between the LMC and SMC Cepheids. We also note the remarkable agreement between the SMC/LMC  P-L slopes for the Period-Wesenheit relations.
\end{enumerate}

\noindent This work is focused on SMC P-L relations, which are found to be linear as opposed to the nonlinear LMC P-L relations reported in our Paper I \citep{nge09} and reference therein. The work of \citet{kan04}, \citet{kan06} and \citet{kan07} has provided one possible theoretical scenario by which the LMC P-L relation can be nonlinear whilst the SMC P-L relation is linear. These relate to metallicity differences leading to different mass-luminosity relations obeyed by Cepheids in the Magellanic Clouds and a different hydrogen ionization front-stellar photosphere interaction in terms of periods and phases. The result of identical P-L slopes in LMC and SMC period-Wesenheit relations is interesting, because the $VI$-band P-L slopes for LMC Cepheids are nonlinear but linear in the case of SMC Cepheids. In terms of period-color relation, \citet{kn04} found that the mean light period-color relations in LMC and SMC are nonlinear and linear, respectively, but \citet{san09} suggested the SMC period-color relation should be nonlinear. To reconcile the linear and identical slopes in LMC and SMC period-Wesenheit relations implies that the $(V-I)$ period-color relation must play a substantial role here. We will address the issue of period-color relations in a future paper.

\acknowledgments

We thank comments from an anonymous referee to improve the manuscript, and H.-J. Kim for assistance with the permutation method. CCN thanks the funding from the Ministry of Science and Technology (of Taiwan) under the contract MOST101-2112-M-008-017-MY3. AB acknowledges the grant 09/045(1296)/2013-EMR-I from Human Resource Development Group (HRDG), which is a division of Council of Scientific and Industrial Research (CSIR), India. This work is supported by the grant provided by Indo-U.S. Science and Technology Forum under the Joint Center for Analysis of Variable Star Data. Part of this work is based on archival data obtained with the {\it Spitzer Space Telescope} and the NASA/IPAC Infrared Science Archive, which is operated by the Jet Propulsion Laboratory, California Institute of Technology under a contract with the National Aeronautics and Space Administration. This publication also makes use of data products from the 2MASS, which is a joint project of the University of Massachusetts and the Infrared Processing and Analysis Center/California Institute of Technology, funded by the National Aeronautics and Space Administration and the National Science Foundation.

\appendix
\section{Correction of $t$-Test Results for LMC Cepheids between OGLE-II and OGLE-III Catalogs in Paper I}

Due to a mistake made in the code for performing the $t$-statistical test, the $p$-value was mis-labelled as expected $t$-value based on a $t$-distribution given $\alpha$ and $\nu$. Therefore, the label ``t'' in Table 2 of Paper I \citep{nge09} should be replaced by ``$p$-value'', and hence the LMC P-L slopes based on OGLE-II and OGLE-III catalogs are all consistent with each other in all bands. The only exception is the P-L slope for Wesenheit function from \citet[][$-3.277\pm0.014$ as compared to $-3.313\pm0.008$ derived in Paper I]{uda99a,uda99c}.

\end{document}